\documentclass[prd,eqsecnum, 
nofootinbib,superscriptaddress]{revtex4-1}
\global\arraycolsep=2pt
\usepackage{stmaryrd}
\usepackage{amsmath}
\usepackage{amssymb}
\usepackage{graphicx}
\usepackage{textcomp}
\usepackage{calrsfs}
\usepackage{yfonts}
\usepackage{bm}

\newcommand{\be}{\begin{equation}}
\newcommand{\ee}{\end{equation}}
\newcommand{\bea}{\begin{eqnarray}}
\newcommand{\eea}{\end{eqnarray}}

\begin{document}
\title{Casimir Energies for Isorefractive or Diaphanous Balls}


\author{Kimball Milton}\email{kmilton@ou.edu}
\affiliation{Homer L. Dodge Department of Physics and Astronomy,
University of Oklahoma, Norman, OK 73019 USA} 
\author{Iver Brevik}\email{iver.h.brevik@ntnu.no}
\affiliation{ Department of Energy and Process Engineering,
Norwegian University of Science and Technology, NO-7491,
Trondheim, Norway}




\begin{abstract}
It is familiar that the Casimir self-energy
of a homogeneous  dielectric ball  is divergent, although
a finite self-energy can be extracted through second order in the deviation
of the permittivity from the vacuum value.  The exception occurs when the
speed of light inside the spherical boundary is the same as that outside,
so the self-energy of a perfectly conducting spherical shell is finite,
as is the energy of a dielectric-diamagnetic sphere with $\varepsilon\mu=1$,
a so-called isorefractive or diaphanous ball.  Here we re-examine that
example, and attempt to extend it to an electromagnetic $\delta$-function
sphere, where the electric and magnetic couplings are equal and opposite.
Unfortunately, although the energy expression is 
superficially ultraviolet finite, additional divergences appear that
 render it difficult to extract a meaningful result in general,
but some limited results are presented.
\end{abstract}



\date{\today}
\maketitle




\section{Introduction}
Although it is clear that Casimir energies between distinct rigid bodies
are finite, even though they arise in a formal way from summation of changes
in the zero-point field energies by material bodies, that finiteness fails for
the self energy of a single body.  (For a detailed review see
\cite{Milton:2010qr}.)  However, for certain special cases, a unique
finite self energy can be extracted.  The classic case is that of the
 perfectly conducting sphere of zero thickness, where
a unique, finite, {\em positive\/} self energy has been extracted by a variety 
of methods \cite{Boyer:1968uf,Balian:1977qr,Milton:1978sf}:
\be
E_B=\frac{0.04617\hbar c}a,
\ee
where $a$ is the radius of the sphere. 

An obvious generalization of a perfecting conducting spherical shell is a
dielectric ball, with a permittivity $\epsilon$ within the spherical volume.
This, however, immediately runs into problems \cite{Milton:1979yx}.  
Although it is possible to identify a finite self-energy in the dilute 
limit, that is, to order $(\epsilon-1)^2$, unremovable divergences
occur in higher order \cite{Brevik:1998zs}.  The weak-coupling limit
coincides with the result obtained by summing the van der Waals forces
between the molecules that make up the medium \cite{Milton:1997ky}:
\be
E_{\rm vdW}=\frac{23\hbar c}{1536\pi a}(\epsilon-1)^2.\label{exdil}
\ee
The divergence in order $(\epsilon-1)^3$ was verified in heat kernel
analyses \cite{Bordag:1998vs}, where the second heat kernel coefficient 
was shown to
be nonzero in that order, resulting in a logarithmic divergence, making
it impossible to extract a finite energy.  Dispersion does not appear
sufficient to resolve this problem.

A possible way out is to consider a ball having both electric permittivity
$\epsilon$ and magnetic permeability $\mu$.  A general statement of
this formulation was given in \cite{Milton:1996wm}, where both the self energy 
and the stress on the sphere were given, consistent with the principle of 
virtual work.  But much earlier Brevik and collaborator realized that
in the special case $\epsilon\mu=1$, that is, when the speed of light is the
same both inside and outside the sphere, the divergences cancel, and
a completely unique finite self energy can be found \cite{Brevik:1982hc,%
Brevik:1982a,Brevik:1984pt,Brevik:1987zi,Brevik:1987zj,Brevik:1997xm}.

Another generalization was explored more recently, that of an electromagnetic
$\delta$-function shell \cite{Parashar:2017sgo}.  This was explored less
completely in \cite{Milton:2010qr}.  In that case there are, in general,
two (transverse) coupling constants, electric and magnetic, and we indicated
there that although in general for finite couplings the self energy was
divergent, in the special case where the two couplings were equal and opposite,
the divergences apparently 
cancel.  In this paper we wish to explore this problem
further.  We will find that 
the modes brought in by the magnetic coupling
  contribute additional  divergences that seem to render
extraction of a finite self energy problematic.  

The outline of this paper is as follows.  In Sec.~\ref{sec2} we re-analyze
the dielectric-diamagnetic ball with the speed of light the same inside
and outside, $\epsilon\mu=1$, and present accurate numerical results which
are slightly better than those given previously.  In Sec.~\ref{sec3}
we will examine the special case of the electromagnetic $\delta$-sphere
where the two coupling are equal and opposite, $\lambda_e=-\lambda_g$,
and carry out the asymptotic analysis to higher order, and identify the
difficulties.  Concluding remarks are offered in Sec.~\ref{concl}.

We use natural units, with $\hbar=c=1$, and Heaviside-Lorentz electromagnetic
units. 
\section{Diaphanous ball}
\label{sec2}
What we shall call a diaphanous ball is spherical volume of radius $a$, in
vacuum, with both electric permittivity $\epsilon$ and magnetic 
permeability $\mu$ such
that $\epsilon\mu=1$, so the speed of light is the same both inside and
outside the sphere is the same.  Here we will ignore dispersion; that was
considered in \cite{Brevik:1987zi}.  The Casimir energy is given by the
formula
\be
E=-\frac1{4\pi a}\int_{-\infty}^\infty dy \, e^{iy\tilde\tau}
\sum_{l=1}^\infty(2l+1)P_l(\cos\delta)x\frac{d}{dx}
\ln\left(1-\xi^2[(e_l(x)s_l(x))']^2\right),\quad x=|y|=|\zeta|a,\label{ediaph}
\ee
where $\zeta$ is the Euclidean frequency, and where
the modified spherical Bessel functions are
\be
s_l(x)=\left(\frac{\pi x}2\right)^{1/2}I_\nu(x),\quad
e_l(x)=\left(\frac{2 x}\pi\right)^{1/2}K_\nu(x),
\ee
with $\nu=l+1/2$.  Here 
\be
\xi=\frac{\mu-\mu'}{\mu+\mu'}=-\frac{\epsilon-
\epsilon'}{\epsilon+\epsilon'},
\ee
 where $\epsilon$, $\epsilon'$ are the exterior and interior values of the
permittivity, and similarly for the permeability.  Note that when
$\xi\to 1$ the familiar expression of the Casimir energy for a perfectly
conducting spherical shell is recovered.  Of particular note are the
point-splitting regulator terms in Eq.~(\ref{ediaph}): $\tilde\tau=\tau/a$ 
is the dimensionless point-splitting parameter in Euclidean time, while
$\delta$ represents point-splitting in the angular (transverse) directions.
The regulator parameters are to be taken to zero at the end of the calculation.

This expression with $\xi=1$ was evaluated accurately first in
\cite{Balian:1977qr,Milton:1978sf}, and has been reconfirmed
several times since \cite{Leseduarte:1996av,Leseduarte:1996ah,%
Nesterenko:1997jr,Lambiase:1998tf}.
In \cite{Brevik:1982hc,Brevik:1982a} the uniform asymptotic expansion (UAE)
for the Bessel functions was used to evaluate the leading term in the
expansion of Eq.~(\ref{ediaph}) for small $\xi$:
\be
E^{(2)}=\frac3{64a}\xi^2.\label{1stuae} 
\ee
(The superscript refers to the order in the UAE, not to the order in $\xi$.)
Some years later, Klich realized that the leading term in the $\xi$ 
expansion could be exactly
computed by using the addition theorem for spherical Bessel functions
\cite{Klich:1999df}:
\be
\sum_{l=0}^\infty (2l+1)s_l(x)e_l(y)P_l(\cos\theta)=\frac{xy}\rho e^{-\rho},
\ee
where $\rho=\sqrt{x^2+y^2-2xy\cos\theta}$.  The exact 
$O(\xi^2)$ result is only about 6\%
larger than the estimate (\ref{1stuae}):
\be
E_2=\frac5{32\pi a}\xi^2.
\ee
As Klich noted, even extrapolating this result to $\xi=1$ gets within 8\%
of the Boyer energy! [The extrapolation of Eq.~(\ref{exdil}) to $\xi=1$ is good
to 2\%.]
(Incidentally, the same exact treatment can be given for the second-order
coefficient for a dilute purely dielectric ball, Eq.~(\ref{exdil}),
first calculated in numerically in \cite{Brevik:1998zs}, but analytically
in \cite{Lambiase:1999zq}, but here, including even the first two terms
in the UAE is still 15\% low.)

In \cite{Brevik:1997xm} the Casimir energy of a diaphanous sphere is 
calculated to higher order in the UAE.  The second-order term in the UAE
is (keeping the $\xi^4$ term, which was not done in \cite{Brevik:1997xm})
\be
E^{(4)}=\frac{9\xi^2}{2^{12}a}\left(\frac{\pi^2}8-1\right)(6-7\xi^2),
\ee
and at $\xi=1$ the sum of $E^{(2)}$ and $E^{(4)}$ is only about 0.5\% high,
while the coefficient of $\xi^2$ in the small $\xi$ expansion is high by about
the same percentage.  This suggests it may be sufficient to remove the first
two terms in the UAE from the logarithm in (\ref{ediaph}) and then add the
the corresponding approximants:
\be
E=E^{(2)}+E^{(4)}+\sum_{l=1}^\infty R_l,\label{rlsum}
\ee
where ($x=\nu z$, $\nu=l+1/2$, $t=(1+z^2)^{-1/2}$)
\begin{eqnarray}
R_l&=&\frac{(2\nu)^2}{4\pi a}\int_0^\infty dz\bigg\{\ln[1-\xi^2(e_l(\nu z)
s_l(\nu z))^{\prime2}]\nonumber\\
&&\quad\mbox{}+\frac{\xi^2t^6}{(2\nu)^2}+\frac{t^6\xi^2}{(2\nu)^4}\left(
\frac{\xi^2}2 t^6-(1-t^2)(2-25t^2+35t^4)\right)\bigg\}.
\end{eqnarray}
Because $R_l$ is finite, the cutoffs can now be dropped.
This equation (\ref{rlsum})
is to be understood as an asymptotic expansion, in that only
some optimal number of terms in the series are to be included.  Only $R_1$
here makes a substantial contribution, as the Table \ref{tab1} shows for 
various values of$\xi$.

\begin{figure}
\includegraphics{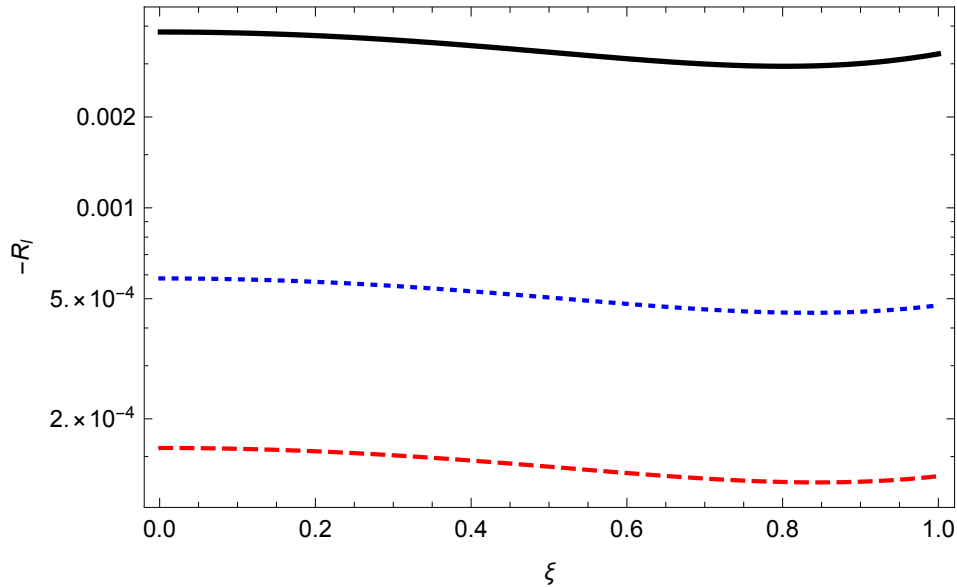}
\caption{\label{fig:rem} The ratio of the remainder contributions to the energy
relative to the lowest order approximant,  $R_l/E^{(2)}$ for isorefractive
dielectric-diamagnetic balls.  Plotted are the negative of this ratio
for $l=1$ (solid, black), $l=2$ (dotted, blue), and $l=3$ (dashed, red).}
\end{figure}

\begin{figure}
\centering
\includegraphics[width=10 cm]{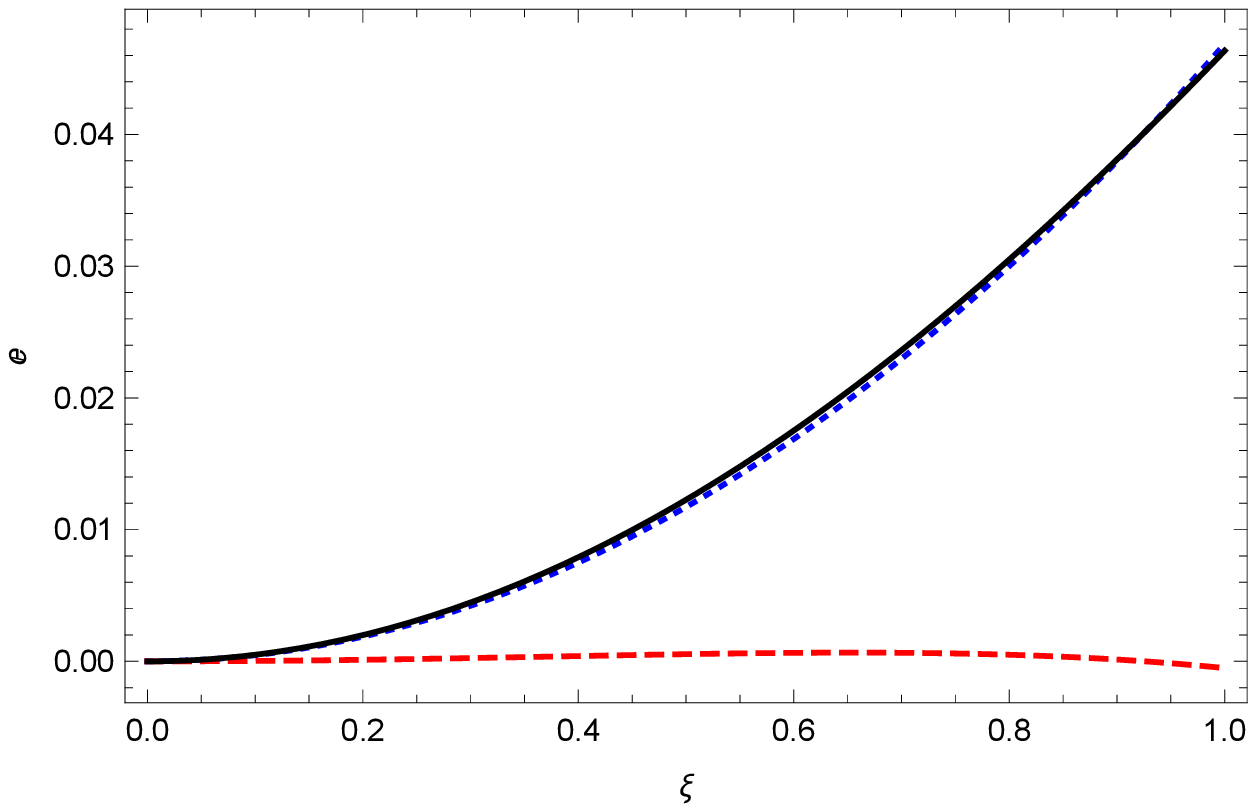}
\caption{\label{fig:e}The energy for isorefractive
dielectric-diamagnetic balls.  Plotted are the first approximation 
 (dotted, blue), the second approximation (dashed, red), and the
total (solid, black).}
\end{figure}

\begin{table}
\caption{\label{tab1}
$E^{(2)}$, $E^{(4)}$, $R_1$, $R_2$, $R_3$, and the sum $E$ for various
values of $\xi$ }
\centering
\begin{tabular}{ccccccc}
\toprule
\textbf{$\xi$}&\textbf{$E^{(2)}a$}	& \textbf{$E^{(4)}a$}& \textbf{$R_1a$}
&\textbf{$R_2a$}&\textbf{$R_3a$}&\textbf{$Ea$}\\
$1$& $0.046875$&$-0.0005135$&$-0.0001517$&$-0.0000223$&
$-6\times 10^{-6}$&$0.04618$\\
$0.5$&$0.011719$&$0.0005456$&$-0.0000384$&$-6\times 10^{-6}$
&$-1.6\times10^{-6}$&$0.0122184$\\
$0.01$		& $4.687\times 10^{-6}$&$3.081\times10^{-7}$&$-1.793
\times10^{-8}$&$-2.74\times 10^{-9}$&$-7.53\times10^{-10}$&$4.974\times10^{-6}$
\\
\end{tabular}
\end{table}
In Fig.~\ref{fig:rem} we show how the remainders rapidly go to zero with
$l$ for all $\xi$.
Since the corrections are so small, for all $\xi$ the lowest UAE contribution
is all that is discernable in a graph of the energy, Fig.~\ref{fig:e}.

Brevik and Kolbenstvedt \cite{Brevik:1982a} give an analytic approximant
to the Casimir energy engineered to be exact for strong coupling $\xi=1$,
but it does not give the exact low-$\xi$ behavior.  The same is true for
the approximation given in \cite{Brevik:1987zj}.  Brevik and Einevoll 
\cite{Brevik:1987zi} include
dispersion in the coupling, but this leads to linearly divergent terms
that are regulated by insertion of an arbitrary parameter.
The direct mode sum given in \cite{Brevik:1997xm} 
includes the first three terms in the UAE,
and is accurate to almost $0.1\%$, 
not quite so accurate as the results reported here.

\section{Dual Electromagnetic $\delta$ Sphere}
\label{sec3}
In general, Casimir
self energies of bodies are divergent, so it is of interest to
study examples where unambiguous finite self energies can be extracted.
Such is the case of a perfectly conducting spherical
shell, or the diaphanous ball
discussed in the previous section, which reduces to the former in the $\xi\to1$
limit.  Another generalization of the spherical shell that would seem to admit
finiteness is an electromagnetic $\delta$-function sphere, with equal and
opposite electric and magnetic couplings; it was observed in 
\cite{Parashar:2017sgo} that
then the catastrophic divergence in third-order in the coupling cancels, 
because only even powers in the coupling appear in the uniform asymptotic
expansion of the energy integrand.  It turns out, however, that this
is a rather more subtle problem than we would have anticipated.

We follow the notation
and formalism given in \cite{Parashar:2017sgo}.  
In this model, the couplings
are modelled by a plasma-like dispersion relation, $\lambda_{e,g}=
\zeta_{p,m}/\zeta^2$, and the form of the Casimir energy, after the bulk
vacuum energy is subtracted, is [analogous to (\ref{ediaph}), except we
have used a more general form of the frequency regulator]
\be
E=-\frac1{4\pi}\sum_{l=1}^\infty (2l+1)P_l(\cos\delta)\int_{-\infty}^\infty
d\zeta\frac{e^{i\zeta\tau}-1}{i\zeta\tau}\zeta\frac{d}{d\zeta}\ln\Delta^E
\Delta^H,
\label{elambdasph}
\ee
where the TE and TM modes are given by
\be
\Delta^{E,H}=1+\zeta^2\frac{\lambda_e\lambda_g}4+|\zeta|[\lambda_{e,g}e_l(x)
s_l(x)-\lambda_{g,e}e_l'(x)s_l'(x)],\quad x=|\zeta|a.
\ee
In \cite{Parashar:2017sgo} we mostly considered the electric case where 
$\lambda_g=0$, but we did remark that interesting cancellations occur if
$\lambda_e=-\lambda_g$.  Here we will explore this further, and define
$\lambda=\zeta_p a=-\zeta_m a$. This leads to the following form for
the quantities in the logarithm:
\be
\Delta^{E,H}=1-\frac{\lambda^2}{4x^2}\pm\frac\lambda{x}[e_l(x)s_l(x)+e_l'(x)
s_l'(x)].\label{isoenergy}
\ee
As we will see in Sec.~\ref{sec-exact},
there is a difficulty with this model,
in that singularities  appear for finite imaginary frequency $\zeta$ 
arising from the $e_l's_l'$
terms. 
\subsection{Uniform Asymptotic Expansion}
 To focus on the ultraviolet behavior, we will modify the
UAE expansions by replacing, as suggested in \cite{Parashar:2017sgo}, $1/z\to
t$, which is correct for large $z$.  Doing so leads to the 
modified UAE expansion (recall $x=\nu z$, $\nu=l+1/2$, $t=(1+z^2)^{-1/2}$)
\begin{eqnarray}
\ln\Delta^E\Delta^H&\sim& -\lambda^2\frac{t^2}{2\nu^2}
-\lambda^4\frac{t^4}{16\nu^4}-\lambda^2\frac{t^6}{192\nu^6}[3(1-6t^2+6t^4)^2
+2\lambda^4]\nonumber\\
&&\quad\mbox{}-\lambda^2\frac{t^8}{512\nu^8}
[2(-1+t)(1+t)(1-6t^2+6t^4)(-13+275t^2-840t^4+630t^6)\nonumber\\
&&\quad\mbox{}+4\lambda^2(1-6t^2+6t^4)^2+\lambda^6]+O(\nu^{-10}).\label{uaed2}
\end{eqnarray}

In \cite{Parashar:2017sgo} we did calculate the energies corresponding to the
first two terms in this uniform expansion, just twice those from the 
$\lambda_e$ contribution. 
The $O(\nu^{-2})$ term in the UAE is in general sensitive to the regulator,
\be
E^{(2)}=\frac{\lambda^2}{4a}\left(1-\frac1\Delta\right),
\ee
where $\Delta=\sqrt{\delta^2+\tilde\tau^2}$.
The divergent term arises from the form of the temporal point splitting in
Eq.~(\ref{elambdasph}).  Had we used a simple 
imaginary exponential instead, and set
$\delta=0$, no divergence would have appeared, as we saw in (\ref{1stuae}).  
However, the form of the
divergence is that expected on general grounds, but it would seem it can be
consistently removed.  The fourth-order term is finite, but
was not explicitly given in \cite{Parashar:2017sgo}. It is twice the $O(\lambda^4)$
term evaluated in Eq.~(4.14) in \cite{Milton:2017ghh}:
\be
E^{(4)}=-\frac{\lambda^4}{16a}\left(\frac{\pi^2}8-1\right).
\ee

\subsection{First Approximation}
\begin{figure}
\includegraphics{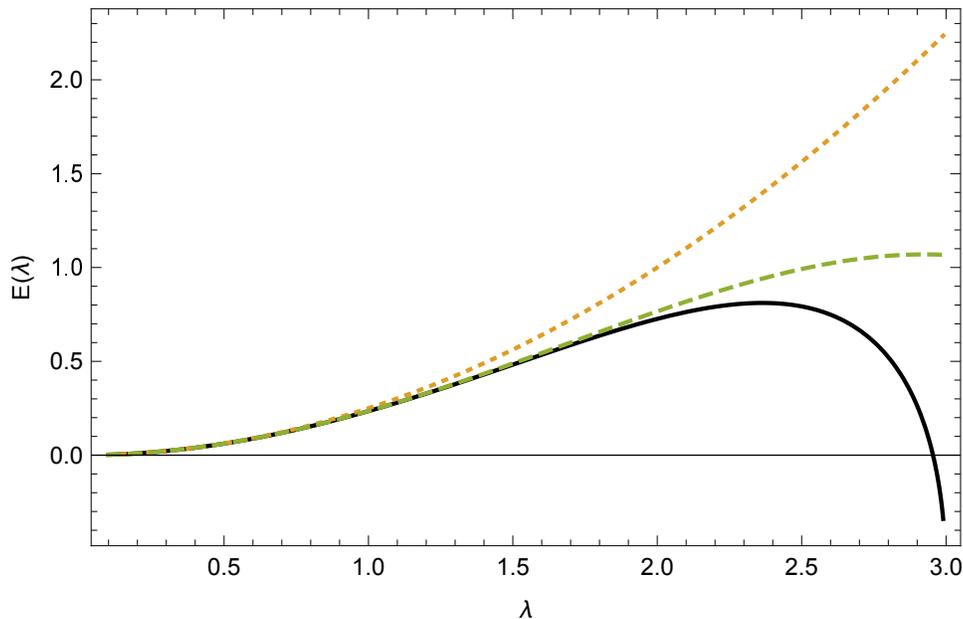}
\caption{\label{fig:leading} Energy estimate $\tilde E$ for a diaphanous ball
(in units of $1/a$) based on the leading terms in the UAE as
a function of the coupling $\lambda$.  Plotted
is the exact approximant (solid curve), 
the contribution of the leading order $E^{(2)}$
(with the $1/\Delta$ divergence removed) (dotted curve), and the sum
of the first two leading orders $E^{(2)}+E^{(4)}$ (dashed curve).
Although the exact approximant is finite at $\lambda=3$, it possesses
infinite slope there, having changed sign for a slightly smaller
value of $\lambda$.}
\end{figure}

Before we consider the general situation, let us see if we can extract
some sort of reasonable approximation to the energy.
We note that the leading terms in Eq.~(\ref{uaed2}) 
(the highest power of $\lambda$ in each order
of $1/\nu$) can be readily summed:
\be
\ln\Delta^E\Delta^H\sim -2\sum_{n=1}^\infty \frac1n 
\left(\frac{\lambda t}{2\nu}\right)^{2n}=2\ln\left(1-\frac{\lambda^2t^2}
{(2\nu)^2}\right).
\ee
The terms we have summed here might be presumed to be the largest
contributions, since for a given power of $\lambda^2$, only the leading
power of $1/\nu^2$ is kept.
Then, if we subtract off the leading, divergent term, we can write
this contribution to the energy as
\be
\tilde E=E^{(2)}+\tilde E_R,
\ee
where
\be
E_R=-\frac{\lambda^2}{\pi a}\sum_{l=1}^\infty f(2\nu/\lambda),\ee
in terms of the function
\be
f(x)=\int_0^1 dt \,t^2\sqrt{1-t^2}\frac1{x^2-t^2}=
-\frac\pi4(1-2x^2+2x\sqrt{x^2-1}),
\ee
which only exists if $x>1$.  For large $x$,
\be
f(x)\sim \frac\pi{16x^2},\quad x\gg1.
\ee
To improve convergence of the $l$ sum, we can subtract the next term
in the UAE, so
\be
\tilde E=E^{(2)}+E^{(4)}-\frac{\lambda^2}{\pi a}\sum_{l=1}^\infty\left[
f\left(\frac{2\nu}\lambda\right)-\frac\pi{16}\left(\frac{\lambda}{2\nu}
\right)^2\right].
\ee
From this, we can readily obtain a numerical estimate, based on this
leading approximation, valid for $\lambda<3$, shown in Fig.~\ref{fig:leading}.
The figure shows the first two UAE contributions, and the sum of all
the leading terms as explained above, with the second-order divergence
removed.  It is to be noted that the exact approximant is finite at
at $\lambda=3$, $\tilde E(\lambda=3)=-0.686434/a$,
but it is singular there, since the derivative becomes
infinite at that point.  Therefore, it is unclear how to analytically
continue this approximant result to higher values of the coupling $\lambda$.

\subsection{$O(\lambda^2)$ Contribution}

We can also, in principle, compute the exact
order-$\lambda^2$ contribution by doing the angular momentum sum exactly again 
using the addition theorem \cite{Klich:1999df}.
Indeed, the $l$-sum over the Bessel functions 
can be thus replaced by
integrals over $w=2x\sqrt{2(1-\cos\delta)}$, for example,
\be
\sum_{l=0}^\infty (2l+1)e_l^2(x)s_l^2(x)=\frac{x^2}2\int_0^{4x}\frac{dw}w
e^{-w}.
\ee
The  divergence at $w=0$  here is irrelevant because of the derivative
appearing in Eq.~(\ref{elambdasph}).  
However, the terms involving derivatives of Bessel functions possess serious
singularities when $\cos\delta=1$, which at present we do not see how to deal 
with.  Although these structures represent the infrared singularities of
the derivatives of the modified Bessel function, because they result in 
divergences when the field-points overlap, they seem to represent ultraviolet
singularities not captured by the UAE.

\subsection{General Analysis}
\label{sec-exact}
\begin{figure}
\includegraphics{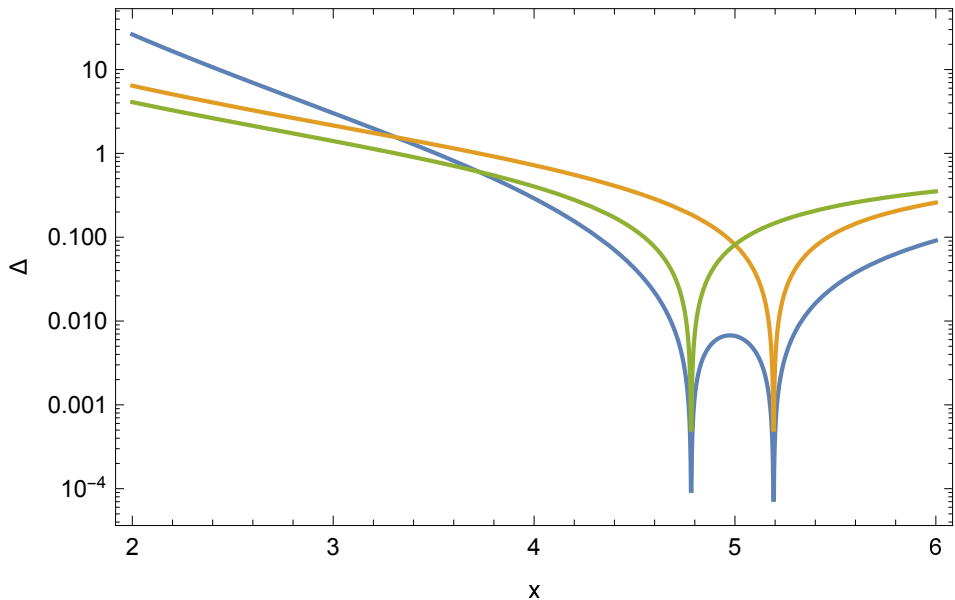}
\caption{\label{fig-zeroes}
Zeroes of $\Delta^E$ (right) and $\Delta^H$ (left), 
and of their product, shown by the plots of their magnitudes, for $l=1$ and
$\lambda=10$.  As $\lambda$ gets large, the zeroes approach $\lambda/2$ for all
$l$.}  
\end{figure}

The difficulties we have encountered in $O(\lambda^2)$ are symptomatic of
a more general pathology.  It is easy to see that $\Delta^E$ always has one 
zero for finite positive $x$, and if $\lambda$ is large enough $\Delta^H$
does as well.  (For $l=1$, the minimum value of $\lambda$ for $\Delta^H$ to
develop a zero is $\lambda_1=8/3=2.666\dots$.)  For large $\lambda$ both zeroes
approach $\lambda/2$, as shown in Fig.~\ref{fig-zeroes}.
These zeroes translate into poles in the frequency integrand in 
(\ref{elambdasph}).
This is rather surprising, since the whole point of doing the 
Euclidean rotation of the frequency to the imaginary axis, $\omega\to i\zeta$,
is to avoid singularities along the real axis.  However, this phenomenon
by itself is not fatal, since we would think the energy would be obtained
by taking the real part of the expression (\ref{elambdasph}),
(\ref{isoenergy}), 
that is, the principal values coming from these simple poles.  But, numerically
this is a bit challenging.

\begin{figure}
\includegraphics{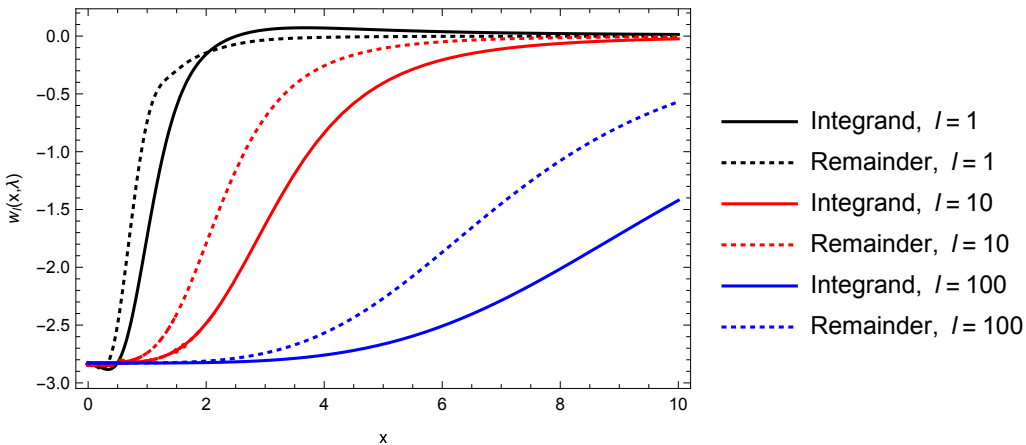}
\caption{\label{integrand}
The integrand of the energy $w_l(x,\lambda)$ given by (\ref{remint})
for $\lambda=1$ and $l=1,10, 100$, shown by the dotted lines.  The solid lines
show the unsubtracted $\ln\Delta^E\Delta^H$ integrand.  The removal of the
leading UAE contributions greatly improves the large $x$ behavior, but leaves 
large and growing contributions for moderate values of $x$.}
\end{figure}

Therefore, a better scheme would seem to be to rotate back through $-\pi/4$,
to a path of integration bisecting the first quadrant of the complex frequency
plane.  Removing the first two approximants coming from the UAE, we then
obtain the following expression, which would seem amenable to numerical
evaluation:
\be
E=E^{(2)}+E^{(4)}
-\frac1{4\pi a}\sum_{l=1}^\infty(2l+1)R_l(\lambda),
\ee
where the remainder is given by (the cutoff has been dropped, in the
expectation that the remainder is finite)
\be
R_l=\int_0^\infty dx\,w_l(x,\lambda),\quad
w_l(x,\lambda)=\Re(1-i)x\frac{d}{dx}
\left[\ln\Delta^E\Delta^h+\frac{\lambda^2t^2}{2\nu^2}+\frac{\lambda^4 t^4}
{(2\nu)^2}\right]_{x\to x(1-i)}.\label{remint}
\ee
The integrand, computed numerically, is plotted in Fig.~\ref{integrand}.
It is seen that subtraction of the leading UAE contributions greatly suppresses
the large $x$ behavior, so the integrals are rapidly convergent.  However, 
because of the small $x$ singularities of the derivatives of the Bessel 
functions, there is a finite contribution at moderate $x$,
all the way down to $x=0$,  which grows with
$l$.  Consequently, when this is integrated over $x$, the remainder $R_l$ grows
with $l$. This growth appears to be cubic. 
 When the $P_l(\cos\delta)$ convergence factor is inserted into the $l$ 
summation, this would translate into a divergence going like $\delta^{-5}$, a
quintic divergence.  So we conclude that the UAE does not capture the real
divergence of the self energy for the isorefractive sphere.
Because of numerical errors, and the likely appearance of a subleading
logarithmic divergence, it appears unfeasible to extract a finite remainder,
even if the divergent terms can be ``renormalized'' away.

\section{Conclusions}
\label{concl}
We have re-examined two situations which extend the classic problem
of the ideal perfectly conducting sphere: the diaphanous dielectric-diamagnetic
ball, where the speed of light is the same on both sides of the spherical
surface, and the isorefractive $\delta$-function sphere, where the electric
and magnetic couplings are equal and opposite.  The former situation has been
well studied, and is uniquely finite; here we extend the accuracy of the 
numerical calculations a bit.  The latter situation, although apparently
ultravioletly finite, possesses infrared sensitivity that translates into
much more severe ultraviolet divergences than revealed by the UAE, 
which seem to make it practically impossible
to extract a well-defined self energy.  This sensitivity manifests itself
as poles on the imaginary frequency axis in the energy integrand.
 By truncating the theory, We are able 
to make some estimates for small coupling $\lambda$.
 Further study of this suprisingly pathological model is warranted.

\vspace{6pt} 



\acknowledgments{This work was supported by grants from the US National
Science Foundation, \#170751, and by the Research Council of Norway, 
\#250346.  We especially want to thank Prachi Parashar for invaluable
collaborative work.}





T




\end{document}